\documentclass[preprint,aps]{revtex4}
\usepackage[toc,page]{appendix}
\usepackage{amsmath,amssymb,graphicx,psfrag}
\usepackage{siunitx}
\renewcommand{\centi}{\ensuremath \mathrm c}

\newcommand{\beq}{\begin{equation}}
\newcommand{\eeq}{\end{equation}}
\newcommand{\bea}{\begin{eqnarray}}
\newcommand{\eea}{\end{eqnarray}}
\newcommand{\nn}{\nonumber}

\newcommand{\la}{\langle}
\newcommand{\ra}{\rangle}
\newcommand{\dg}{\dagger}

\newcommand{\f}{\frac}
\newcommand{\lbr}{\left(}
\newcommand{\rbr}{\right)}
\newcommand{\lbs}{\left[}
\newcommand{\rbs}{\right]}

\newcommand{\empformula}[1]{\ensuremath \mathrm{#1}}
\newcounter{subfigure}[figure]
\alph{subfigure}

\begin{document}
\title{Universal features of Thermopower in High $T_c$ systems and Quantum Criticality}

\author{ Arti Garg, B Sriram Shastry}
\affiliation{Physics Department, University of California, Santa Cruz, California 95064, USA}
\author{Kiaran B. Dave, and Philip Phillips}
\affiliation{Physics Department, University of Illinois, Urbana Champaign,  IL 61801}

\begin{abstract}
In high $T_c$ superconductors a wide ranging connection between the doping dependence of the transition temperature $T_c$ and the room temperature thermopower $Q$ has been observed.  A ``universal correlation'' between these two quantities exists with the thermopower vanishing at optimum doping as noted by OCTHH (Obertelli, Cooper, Tallon, Honma and Hor). In this work we provide an interpretation of this OCTHH universality in terms of a possible underlying quantum critical point (QCP) at $T_c$. Central to our viewpoint is the recently noted Kelvin formula relating the thermopower to the density derivative of the entropy. Perspective on this formula is gained through a model calculation of the various Kubo formulas in an exactly solved 1-dimensional model with various limiting procedures of wave vector and frequency.
\end{abstract}


\maketitle
\section{Introduction}
Universal properties of strongly correlated matter are particularly
interesting, since they hold the promise of revealing the fundamental
physics of these systems. An example in the case of heavy fermion
systems is the well-known Kadowaki--Woods
relation~\cite{kadowaki-woods}  $A\;\gamma^{-2}\simeq1.0\times
10^{{-5}} \ \micro \ohm \;\centi \meter $ between the specific heat
coefficient $\gamma$ and $A$, the $T^{2}$ coefficient of the
resistivity. Understanding the origin of this universal value has led
to considerable theoretical progress~\cite{heavy-fermi}. In the case of High $T_{c}$ materials, Obertelli, Cooper and Tallon
(OCT)~\cite{Tallon} observed that the thermopower for several cuprates vanishes in the vicinity of optimal doping. Honma and Hor (HH)~\cite{honmahor} extended their analysis, which was based on the phenomenological scale $1-\frac{T_c}{T_c^{\rm max}}=82.6(x-0.16)^2$~\cite{presland},  by first showing that the thermopower for all the cuprates
collapses onto a universal curve as a function of the doping level
with a zero crossing at a hole content of $x\approx 0.23$.  Since this
universal scale is valid for all the cuprates, they then advocated~\cite{honmahor} that it
can be used as an independent calibration of the doping level.  Consequently, the mapped out the phase diagram
of the cuprates in which $x$ was determined from the thermopower not the Presland \cite{presland} scale.  Using
this scale, they showed~\cite{honmahor} that optimal
doping for 19 of the 23 cuprates studied corresponded with the zero crossing of the thermopower. The exceptions are 4
single-layer materials. Assuming this is not a coincidence, it is reasonable to conclude that the mechanism of high-T$_c$ and
the vanishing of the thermopower~\cite{false-hf} share a common origin.

The observation of OCTHH has stimulated considerable thought in the
community~\cite{leggett,phillips,shila,Shastry}. In the context of the
Hubbard model, some~\cite{phillips,shila} have argued that dynamical
spectral weight transfer ceases at the doping at
which the thermopower vanishes, thereby defining a quantum phase transition
(QPT) where the upper-Hubbard band decouples and Fermi-liquid theory obtains. While cluster calculations on the Hubbard
model~\cite{jarrell} support this interpretation, no simple model has been
studied in which exact statements can be made regarding the vanishing
of the thermopower and the onset of a quantum phase transition. The present work is stimulated by this situation, and we present below two key ideas and a set of model calculations that provide 
a natural framework for understanding such a universality. 

\begin{figure}[h!]
\begin{center}
\includegraphics[width=2.0in,angle=-90]{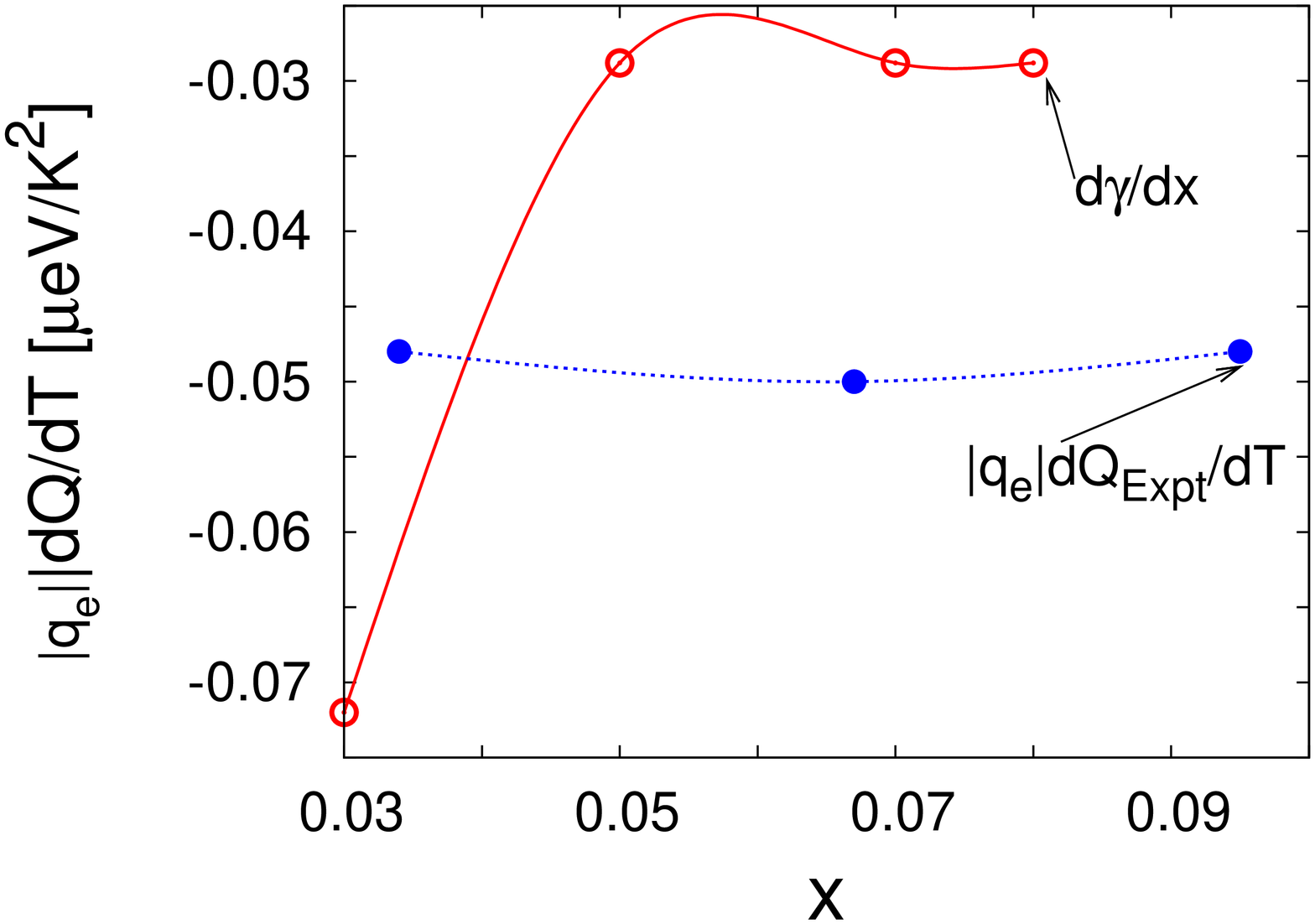}
\includegraphics[width=2.0in,angle=-90]{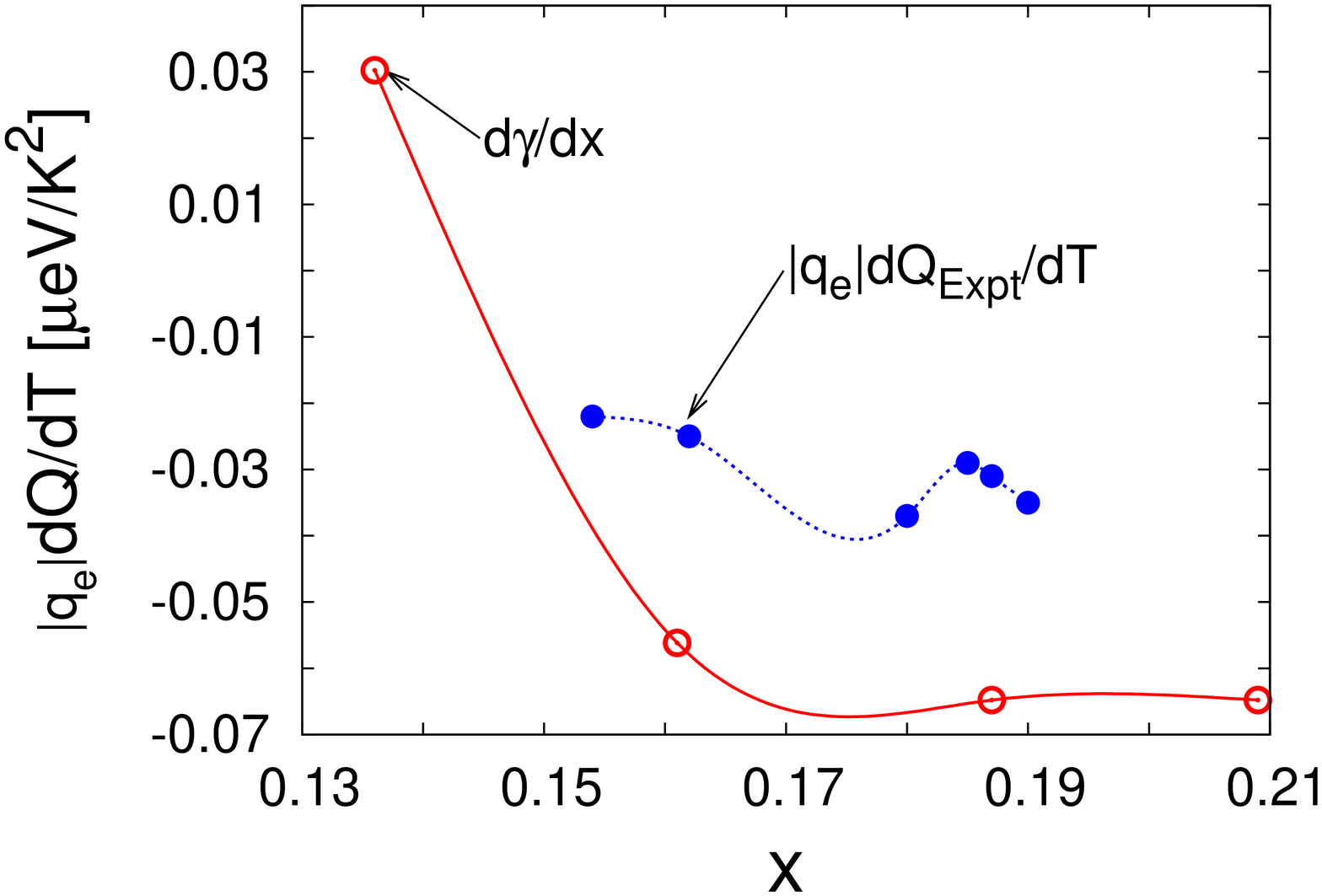}
\vskip 0.5cm
\caption{A test of the reliability of the Kelvin formula, as explained in the text, for $\empformula{Tl_2Ba_2CuO_{6+\delta}}$ at $T=100\;\kelvin $ (left) and BSSCO at $T=120\;\kelvin$ (right). The red solid curves are $d\gamma/dx$ and the blue dotted curves are $|q_e| \ dQ/dT$.  These curves would coincide in the $T\rightarrow 0$ limit if the Kelvin formula were exact .}
\label{fig:cqdata}
\end{center}
\end{figure}

The first key idea in our work is the interpretation of the thermopower as being largely determined by {\em thermodynamics}, rather than transport aspects, such as velocities and relaxation times \footnote{ This separation of the thermopower into the two components of thermodynamics and transport is well illustrated by rewriting the Mott formula for the thermopower of a weakly diffusive metallic system given in textbooks~\cite{Ashcroft-Mermin} as:
$$
Q_{\mbox{Mott}}= T \frac{\pi^{2}k_{B}^{2}}{3 q_{e}} \ \{  \frac{d}{d \mu}  \ln[  \rho(\mu)] + \frac{d}{d \mu}  \ln[ \langle (v^{x}_{p})^{2} \tau(p,\mu)] \}, 
$$ where $\mu$ is the chemical potential. In this expression, the first term gives the density of states (and hence thermodynamic) contribution, and the second term gives the transport contribution from the Fermi-surface average of the squared velocity and the relaxation time.
}. While this idea of thermodynamic domination of thermopower cannot
be an exact statement, it leads to the Kelvin formula proposed by Shastry and coworkers~\cite{Shastry,Shastry-review} in the spirit of Lord Kelvin's original treatment~\cite{kelvin}. The Kelvin formula for thermopower $Q_{K}$ is obtained by computing the slow limit of an exact formula at finite $q,\omega$, and is given by \footnote{Throughout this paper, the slow limit denotes taking $\omega\rightarrow 0$ followed by $q\rightarrow 0$, whereas the fast limit denotes $q\rightarrow 0$ followed by $\omega\rightarrow 0$. See J. M. Luttinger, Phys. Rev. {\bf135}, A1505 (1964).}
\beq
Q_{K} = \f{1}{q_e }\lbr \f{\partial S}{\partial x}\rbr_{T,V}, \label{kelvin}
\eeq
where $S$ is the entropy density, $x$ is the density of carriers in the system and $q_e$ is the charge of the carriers ($-|q_e|$ for electrons)\footnote{For applying this formula to holes, we regard $x$ as the hole  density and remember to use $q_{e}=-|q_{e}|$. }. In brief, this approximate formula for thermopower  captures the enhancements due to all fluctuations that influence the thermodynamics of a many body system. While theoretical benchmarks of this approximation exist~\cite{Shastry}, it is also useful to check its consequences directly for the high-$T_{c}$ systems.  Using standard thermodynamics, one gets the following relation between the temperature dependence of the Kelvin thermopower and the specific heat variation with particle density $x$:
\beq
q_e\f{\partial Q_{K}}{\partial T} = \frac{1}{T} \lbr \f{\partial C_v}{\partial x}\rbr_{V,T}\underset{T\rightarrow 0}{\longrightarrow} \lbr \f{\partial \gamma}{\partial x}\rbr_{V,T} ,
\label{kelvin2}
\eeq
 where $ \gamma $ is the low-temperature coefficient of the specific heat. This equality comprises a relationship between  two independent experiments and can therefore be tested with experimental data. Fig.~\ref{fig:cqdata} plots the left and right side of (2) based on thermopower~\cite{Tallon} and electronic specific heat data~\cite{Loram1,Loram2} for $\empformula{Bi_2Sr_2CaCu_2O_{8+\delta}}$ and $\empformula{Tl_2Ba_2CuO_{6+\delta}}$.
While the agreement is far from perfect  and deviations are certainly expected since the correspondence between the left and right-hand sides of Eq. (\ref{kelvin2}) is only expected to hold at $T=0$, the signs and orders of magnitude of each side of (\ref{kelvin2})
are compliant with the data, suggesting the level of accuracy one can expect from the Kelvin formula in these systems.  

The second key idea involves an underlying quantum critical point~\cite{shila,jarrell} (QCP) as the origin of  the OCTHH universality. We suggest  that  optimum doping corresponds to a QCP at $T=0$, so that at room temperature,  the  entropy  has a  maximum as a function of the density. From the Kelvin formula, this feature would  explain why  the room temperature thermopower changes sign at optimal doping. 

These ideas may be illustrated within a simple fermionic model exhibiting a QCP. The reader is forwarned that this model is quite unphysical because the number of particles is not conserved; nevertheless, its exact solvability makes it invaluable for the purpose at hand. It also has the essential feature that the parameter tuned to reach the QCP is thermodynamically conjugate to the average particle number density. Because the fingerprints of the $T=0$ transition at finite temperature include a  maximum in the entropy as a function of the density, the properties of the model lead naturally to the OCTHH universality.
 
\section{Model} 
We analyse the anisotropic quantum XY model in the presence of a transverse field in one dimension, which, after a Jordan--Wigner transformation, is described by the fermionic  Hamiltonian~\cite{Katsura}${}^{,}$\footnote{In Katsura's~\cite{Katsura} expressions, we  use the notation $J= \frac{1}{2}(J_x+J_y)$ and $\Gamma= \frac{1}{2J}(J_x-J_y)$, where $\Gamma$ is a dimensionless anisotropy parameter.}, 
\beq
\f{H}{J}=- \sum_i \lbs c^\dg_ic_{i+1} + \Gamma c^\dg_ic^\dg_{i+1} + h.c.\rbs -h\sum_i (1-2c^\dg_i c_i).\label{Hamiltonian}
\eeq
  Here $- 2 h$ is the dimensionless  chemical potential.  Due to the $\Gamma$ terms, the number of fermions is not conserved and hence only the average particle number may be fixed. The Hamiltionian can be diagonalized via the Bogoliubov transformation: its eigenvalues are
\beq
\epsilon_k = \pm 2 \sqrt{(h-\cos{k})^2+\Gamma^2\sin{k}^2}\mbox{,} \quad \  -\pi < k\le \pi.
\eeq
For any value of $\Gamma$, the system has a QCP at $h=1$,  where the spectral gap $\Delta = 2|(1-h)|$ vanishes continuously. The $T$--$h$ phase diagram is shown in Figure 2a. There are two low-$T$ phases.  For $h < 1$, the low-$T$ phase has  an energy  gap with a  high density  of fermions.  At  $T=0, \  \Gamma >0 $ and with  $\lvert h\rvert\leq 1$, the equal-time spin correlation function $C(R,t=0)= \la \sigma_i^x \sigma_{i+R}^x \ra$\footnote{As per the Jordan-Wigner transformation, $\sigma^x_i\sigma^x_{i+R}=(c_i+c^{\dagger}_i)\prod_{i\leq j<i+R} (1-2n_j)(c_{i+R}+c^{\dagger}_{i+R})$.} in the limit $R\rightarrow \infty$ is non-zero, being equal to \cite{Mccoy} $C(\infty,0) = \f{1}{2(1+\Gamma)}\lbs \Gamma^2(1-h^2)\rbs ^{\f{1}{4}}$. This result indicates the presence of magnetic long-range order in the ground state. For $h>1$, the low-$T$ phase has a spectral  gap and with a low number density. It corresponds to a quantum paramagnet in the original spin model.

\begin{figure}[h!]
\begin{center}
\includegraphics[width=2.0in,angle=-90]{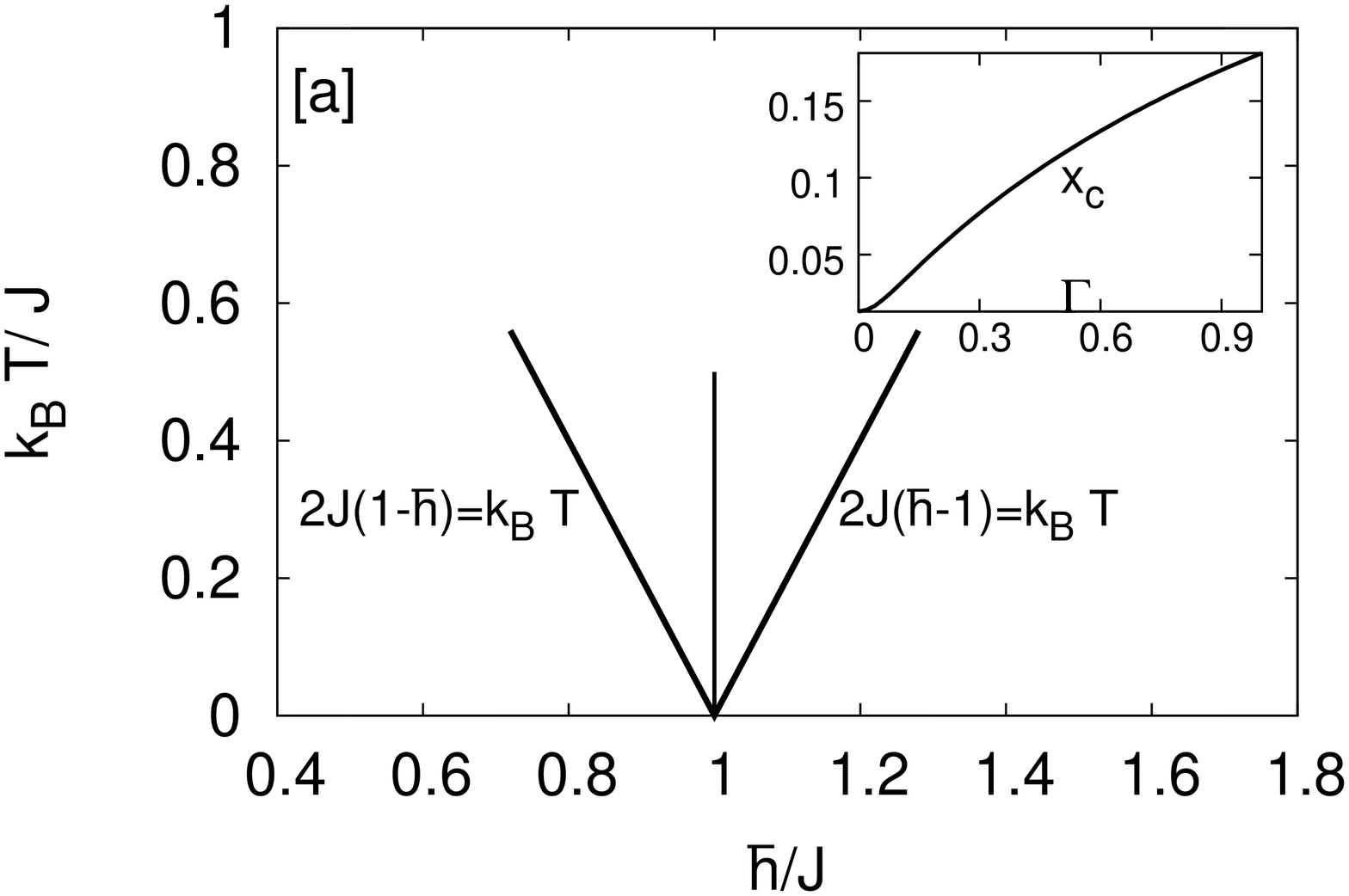}
\includegraphics[width=2.0in,angle=-90]{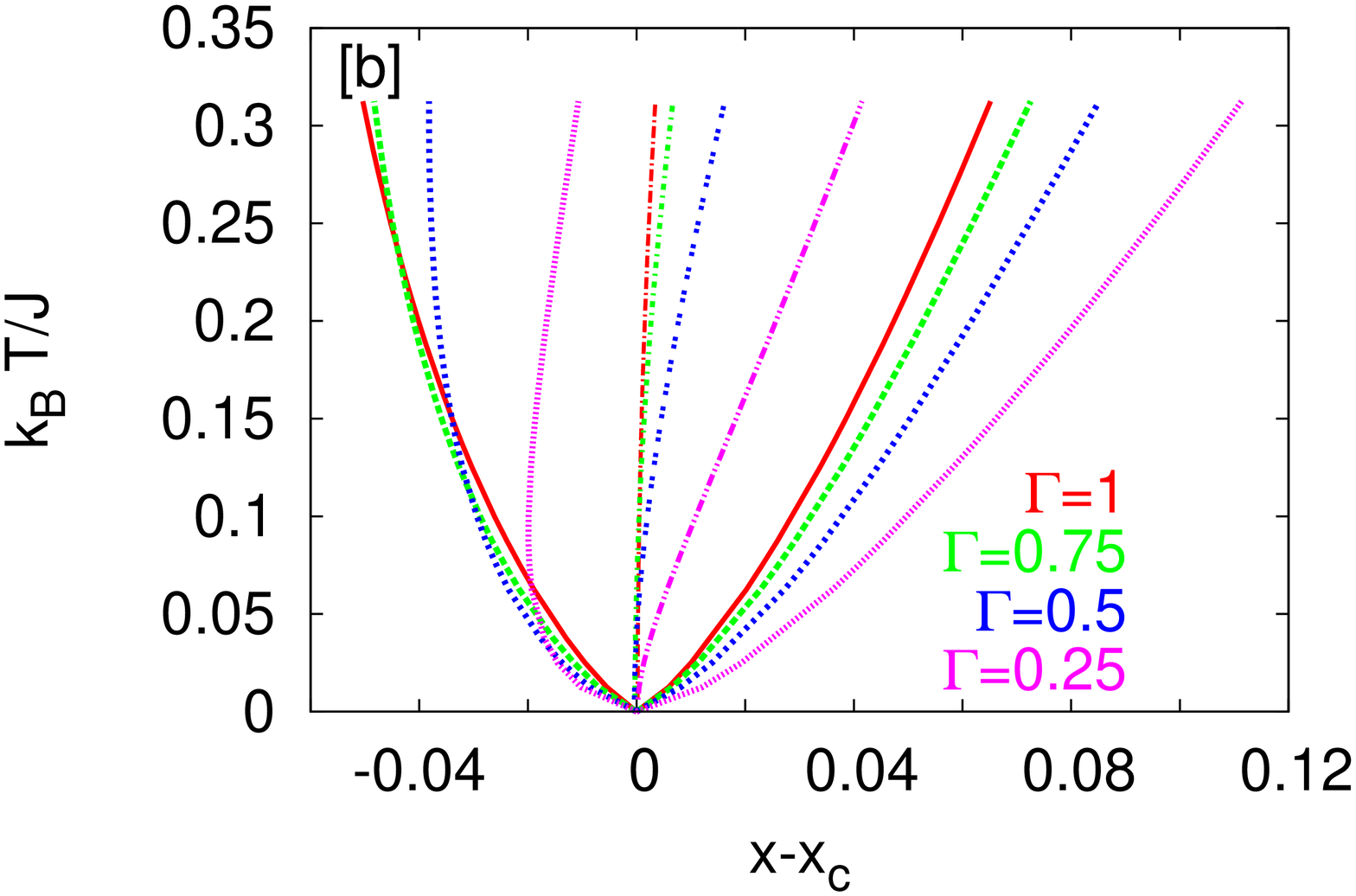}
\vskip 0.5cm
\caption{a) Phase diagram of the 1D quantum model described by the Hamiltonian (\ref{Hamiltonian}) as a function of the chemical potential $h$ and temperature $t$. There is a quantum phase transition at $t=0$ and $h=1$  for any value of anisotropy parameter $\Gamma$. The entropy is a maximum along the central vertical line that we call the``peak line''. The two lines $\Delta=t$  mark the crossovers from the  low-$T$ gapped phases to a high-$T$ phase. Inset: The particle density at the critical point $x_c=x(h=1,T=0)$, which varies significantly as a function of the anisotropy $\Gamma$. b) $t$--$(x-x_c)$ phase diagram of the same model for various values of the anisotropy parameter $\Gamma$. The peak line and the crossover lines in $t$--$h$ plane split into many lines in the $t$--$x$ plane, depending on the value of $\Gamma$.}
\label{xy}
\end{center}
\end{figure}

The lines $\Delta=t$ shown in Fig.~\ref{xy}a have vanishing excitation energies and represent the crossover  between the low-$T$ gapped phases and the intermediate phase, where the dimensionless temperature is defined as $t= k_B T/J$.
The average particle density $x=\la c^\dg_ic_i \ra$ is 
\beq
x=\f{1}{2}\lbs 1-\f{2}{\pi}\int_0^{\pi}~dk~ \tanh{(\f{|\epsilon(k)|}{2t})} \ \f{(h-\cos{k})}{|\epsilon(k)|}\rbs. \label{map}
\eeq
The critical particle density $x_c(\Gamma)=x(h=1,\Gamma)$ at $T=0$ is plotted in the inset of Fig. \ref{xy}a. Due to the nontrivial mapping between $h$ and $x$ in (\ref{map}), the $t$--$x$ phase diagram is dependent on the magnitude of  $\Gamma$, as is shown in Fig. \ref{xy}b. On the other hand, the $t$--$h$ phase diagram is independent  of $\Gamma$. The lines corresponding to a vanishing gap at $h=1$ bend as one moves along the $t$ axis. The bending angle varies with $\Gamma$ and is larger for smaller values of $\Gamma$. In addition, the crossovers $\Delta=t$ between the low-$T$ and high-$T$ phases move as $\Gamma$ is varied. 

\begin{figure}[h!]
\begin{center}
\includegraphics[width=2.0in,angle=-90]{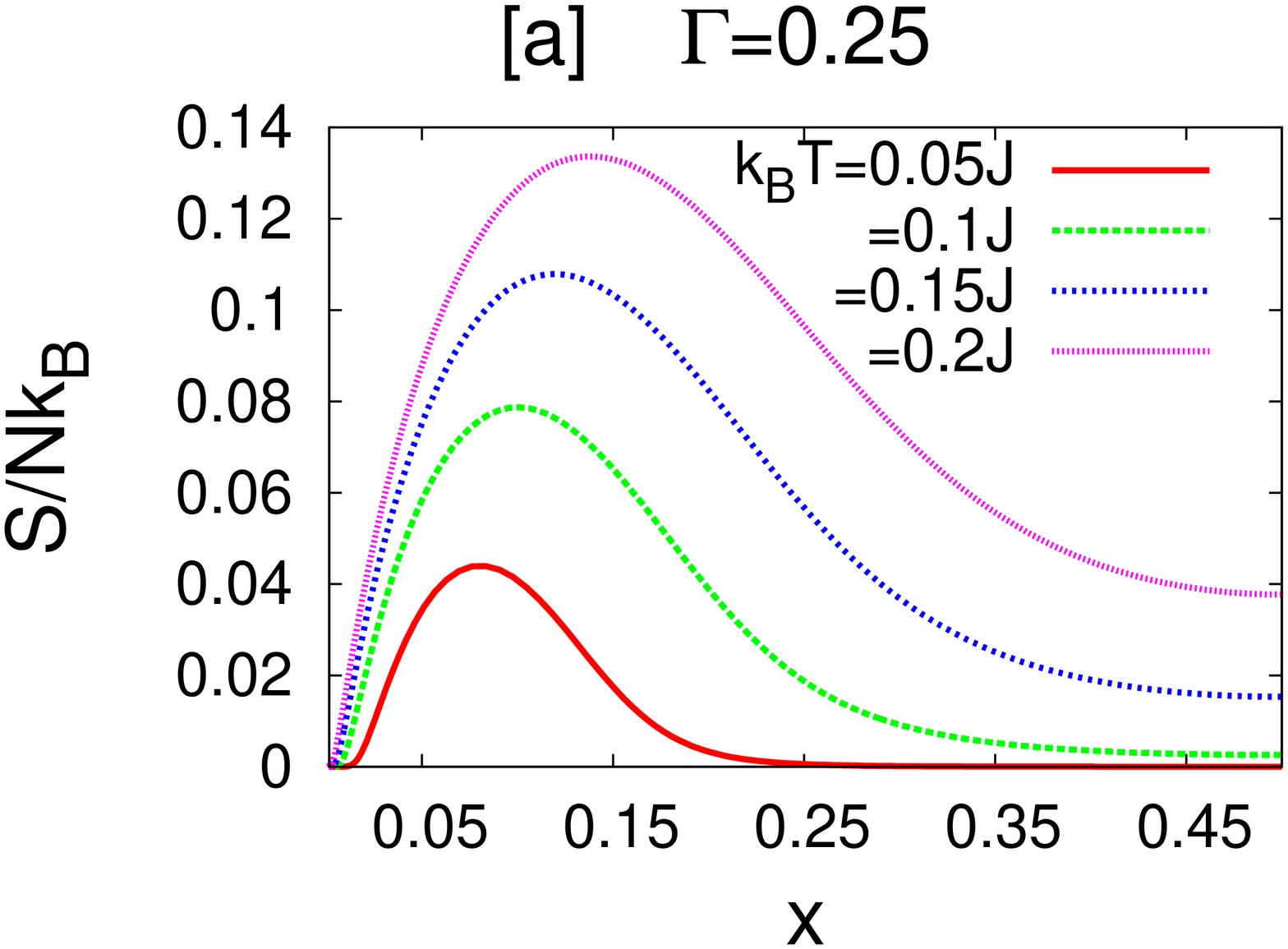}
\includegraphics[width=2.0in,angle=-90]{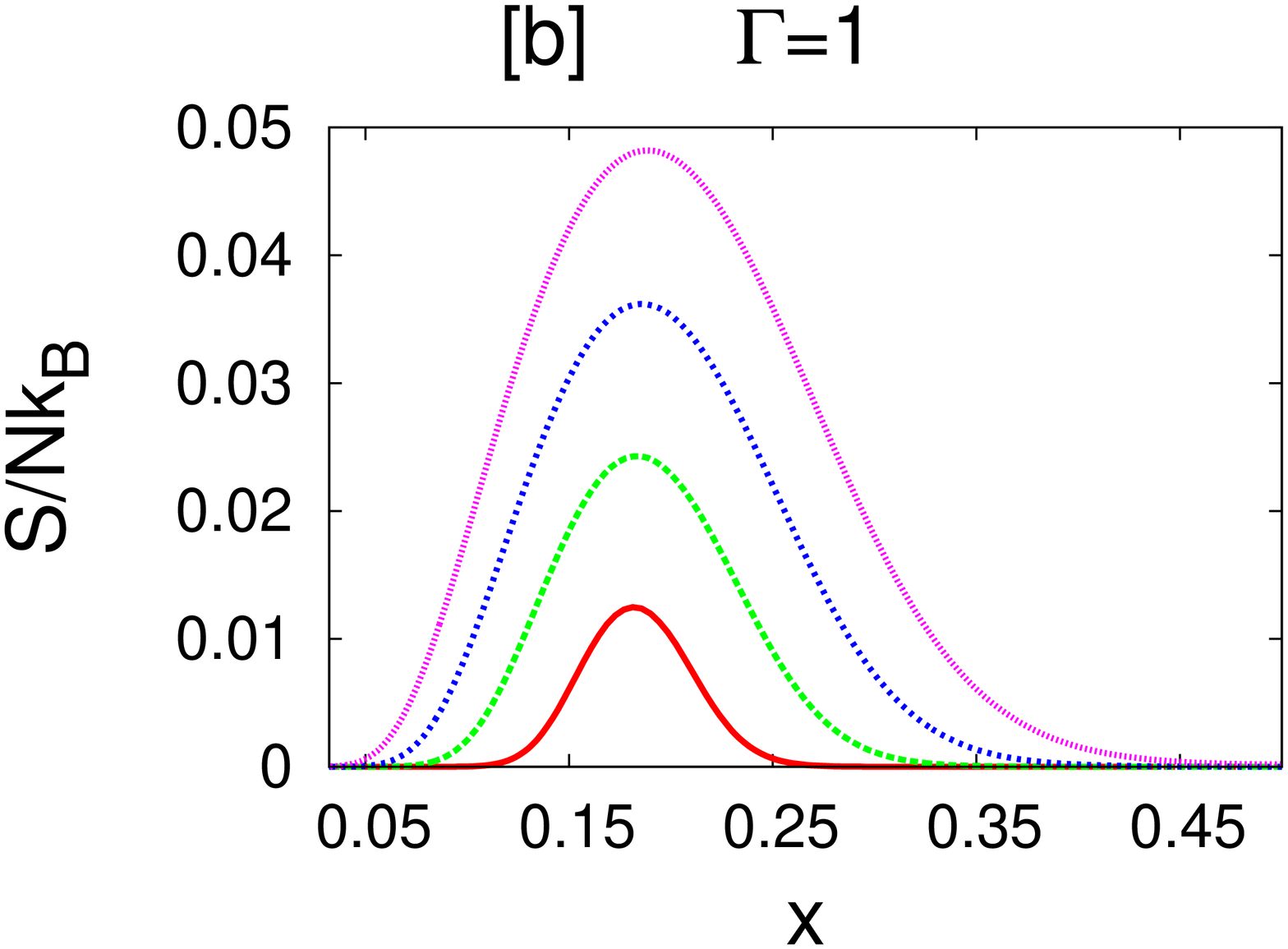}
\caption{Plots of the entropy $S$ as a function of hole density $x$ at several values of $t$ for a) $\Gamma=0.25$ and b) $\Gamma=1$.  Notice the clear shift in the location of the maximum of $S$ as $t$ increases for $\Gamma=0.25$. This corresponds to the ``peak line'' shown in Fig.~\ref{xy}b.}
\label{S}
\end{center}
\end{figure}

 The entropy $S$ for this model may also be found exactly~\cite{Katsura}. For all $t$, $S$  is a maximum as a function of density at $x(h=1)$, as shown in Fig.~\ref{S}. The location of the maximum  has its origins in the large number of micro-states possible for the original spin system at $h=1$. The locus of the maxima of $S$ depends upon $\Gamma$ and is depicted as the central line in Fig.~\ref{xy}b. We term this  the``peak line''. The  peak line has not received attention in earlier works but plays an important role for our purpose. 

\subsection{Choice of current operators}
Eq. (\ref{Hamiltonian}) is the spin-less version of the 1D BCS reduced Hamiltonian. As the total number of particles is not conserved, the standard continuity equation for the charge density $\rho(i)$ does not hold and there is some ambiguity as to the choice of local charge current operator. We work with the standard charge current operator given by $J(n) = i   ( c^{\dg}_nc_{n+1}-c^{\dg}_{n+1}c_n) $ (here $q_e=1$). With this choice, the continuity equation for the charge density has pair sources and sinks as in the BCS problem~\cite{Nambu}. 
The continuity equation for the local energy density is standard $\f{\partial H(n)}{\partial t} +(J_{E\;(n+1)}-J_{E\;(n)}) =0$ with energy current operator given by $J_{E\;(n)}= i \ (1-\Gamma^2) \  \left(c^\dagger_{n-1}c_{n+1}-c^\dagger_{n+1}c_{n-1}\right) + i \ 2 h \left(c^\dagger_nc_{n-1}-c^\dagger_{n-1}c_n\right) +  i \ 2 h \Gamma \left(c^\dagger_{n-1}c^\dagger_n- c_nc_{n-1}\right)$. It is interesting that both the charge and the energy currents are conserved in this integrable model. However, the calculations do not simplify on account of this feature and so we will not pursue its consequences further. 

\subsection{Formulae for the thermopower} For a generic 1D system there are four possible linear response formulae for the momentum- and frequency-dependent thermopower:
\begin{subequations}
\bea
T \ Q_1(q,\omega)& = & \frac{\chi_{\rho(q),H^\dagger(q)}(\omega)}{\chi_{\rho(q),\rho^\dagger(q)}(\omega)}\label{kelvinformula} \\
T \ Q_2(q,\omega) & = & \frac{\chi_{J(q),H^\dagger(q)}(\omega)}{\chi_{J(q),\rho^\dagger(q)}(\omega)}\label{kuboformulas}\\
 T \ Q_3(q,\omega) & = & \frac{\chi_{\rho(q),J_E^\dagger(q)}(\omega)}{\chi_{\rho(q),J^\dagger(q)}(\omega)}\label{spurious1} \\
 T \ Q_4(q,\omega) & = & \frac{\chi_{J(q),J_E^\dagger(q)}(\omega)}{\chi_{J(q),J^\dagger(q)}(\omega)}\label{spurious2}
\eea
\end{subequations}
Here $A(q) = \sum_n e^{-iq\cdot n}A(n)$ denotes the Fourier transform of the local operator $A(n)$.
It can be shown by integration by parts that these four formulas are equivalent if $\left[\rho(0),H(0)\right]=0$, i.e., particle number is conserved. The fast limit of (\ref{kuboformulas}) yields the celebrated Kubo formula; the slow limit of (\ref{kelvinformula}) is (\ref{kelvin}) the Kelvin formula~\cite{Shastry-review}.

\begin{figure}[h!]
\begin{center}
\includegraphics[width=4.0in,angle=0]{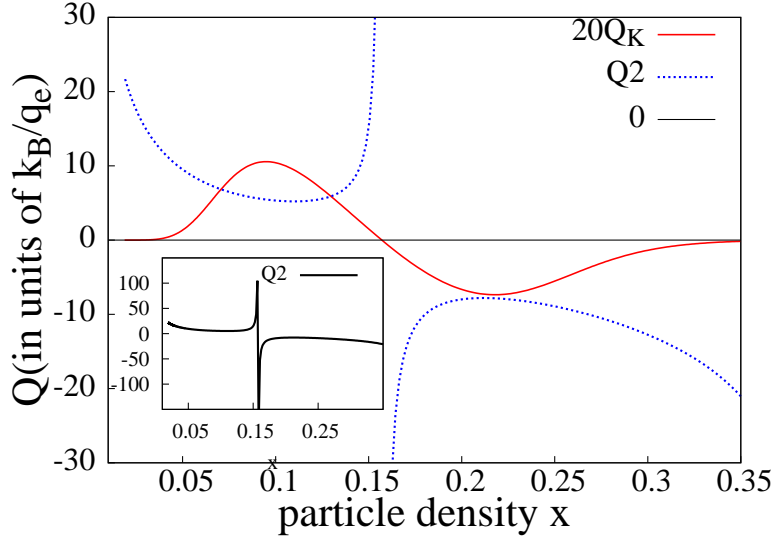}
\caption{The thermopower, as calculated from the Kubo formula (blue dotted curve) and the Kelvin formula (red solid curve). In both the cases, the thermopower is positive for $x < x_c$ and negative for $x>x_c$. The result shown is for $t=0.1$ and $\Gamma=0.75$.}
\label{Q}
\end{center}
\vskip-6mm
\end{figure}

The thermopower was calculated in the slow and fast limits for all four linear response expressions. The Kubo formula is given by
\beq
Q_{Kubo} = Q_2^{\mbox{fast}}= \frac{1}{ T }\frac{\sum_k \sin(k)\f{\partial n_{k}}{\partial k} F(k,h)} { \sum_k \sin(k)A_1(k)\f{\partial n_{k}}{\partial k} }, \label{Qkubo}
\eeq
where $F(k,h)= 2[(h-\cos k) A_1(k)+\Gamma \sin(k)B_1(k)]$,  $A_1(k) = \f{2(h-\cos(k))}{\epsilon(k)}$, $B_1(k)= \f{2\Gamma \sin(k)}{\epsilon(k)}$, and $n_k$ is the Fermi function.
Fig.~\ref{Q} plots the thermopower calculated from the Kubo and the Kelvin formula for $t=0.1$ and $\Gamma=0.75$.
Note that the Kubo formula result changes sign across the QCP through a  divergence. It is positive for $x < x_c$ and negative for $x>x_c$. The Kelvin formula captures the broad features of the thermopower from the Kubo formula result, but it instead changes sign through a zero.

\begin{figure}[h!]
\begin{center}
\includegraphics[width=4.0in,angle=0]{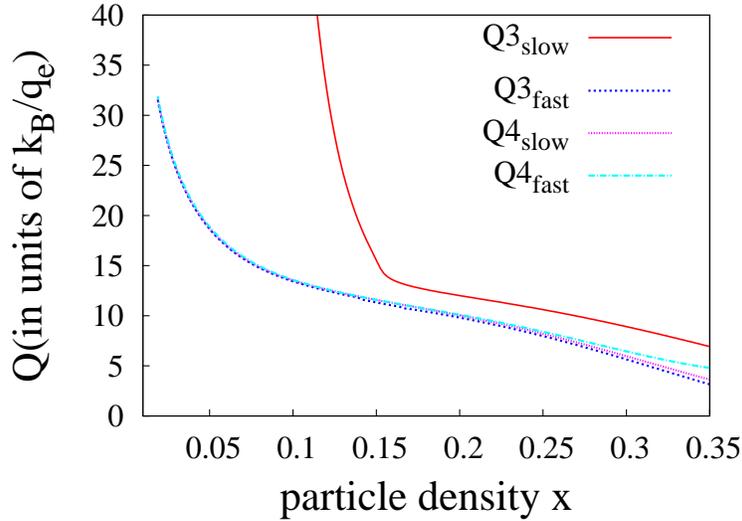}
\caption{The thermopower, as calculated from the formulae  (\ref{Q34}). Both the slow and the fast limits of $Q_{3,4}$, which give the thermopower in the presence of a transverse electromagnetic field, are positive for all values of the particle density $x$ and show no sign change across the QCP. The plot shown is for $t=0.1$ and $\Gamma=0.75$.}
\label{Q2}
\end{center}
\vskip-6mm
\end{figure}

The fast limit of $Q_1$ and the slow limit of $Q_2$ are $O(q)$ and hence vanish. Aside from those deriving from $Q_1$ and $Q_2$, there are four more possible expressions for the thermopower which come from the slow and the fast limit of $Q_3$ and $Q_4$:
\bea
T Q_3^{\mbox{slow}} & =& \f{\sum_k V_H(k)\lbr\f{2\sin k}{\epsilon(k)}+\f{2(h-\cos k)\epsilon^\prime(k)}{\epsilon^2(k)}\rbr \f{1-2n(k)}{2\epsilon(k)}}{\sum_k V(k) \lbr\f{2\sin k}{\epsilon(k)}+\f{2(h-\cos k)\epsilon^\prime(k)}{\epsilon^2(k)}\rbr \f{1-2n(k)}{2\epsilon(k)}} \nn \\
T Q_3^{\mbox{fast}} &= &\frac{\sum_k V_H(k) A_1(k)\f{\partial n_{k}}{\partial k}}{\sum_k V(k) A_1(k)\f{\partial n_{k}}{\partial k}}\nn \\
T Q_4^{\mbox{slow}} & = & \f{\sum_k V_H(k)V(k)\f{\partial n_{k}}{\partial \epsilon(k)}}{\sum_k V^2(k) \f{\partial n_{k}}{\partial \epsilon(k)}} \nn \\
T Q_4^{\mbox{fast}} & =&   \f{\sum_k V_H(k)V(k)(\epsilon^\prime(k))^2\f{\partial n_{k}}{\partial \epsilon(k)}}{\sum_k V^2(k)(\epsilon^\prime(k))^2 \f{\partial n_{k}}{\partial \epsilon(k)}} 
\label{Q34}
\eea 
Here $V(k)= 2\sin(k)$, $V_H(k)=2h V(k)-2(1-\Gamma^2)\sin(2k)$ and $\epsilon^\prime(k)=\f{\partial \epsilon(k)}{\partial k}$.
Fig.~\ref{Q2}  shows plots of the formulae (\ref{Q34}). These four expressions for the thermopower give very similar results: they are positive for all values of the particle density $x$ and do not show any sign change across the QCP.

In the end, one must make a decision as to which thermopower is definitive. On the basis of linear response theory, one knows that formulae derived from (\ref{kelvinformula}) and (\ref{kuboformulas})  concern the response of the system to a longitudinal electromagnetic field while formulas (\ref{spurious1}) and (\ref{spurious2}) concern a transverse electromagnetic field. Since the thermopower measured experimentally is actually the response to a longitudinal electromagnetic field, we believe that only results from (\ref{kelvinformula}) and (\ref{kuboformulas}) contain information about the physical thermopower
 \footnote{At this stage it might be useful to notice the difference between this model and the BCS problem. In the BCS problem, one get accurate response to the transverse electromagnetic field even in simple pairing approximation. But to get the correct response to the longitudinal electromagnetic field, one needs to go beyond the mean field pairing picture and keep collective density fluctuation mode. But in this model there is no possibility like this. Thus we will work with formulas (\ref{kelvinformula}) and (\ref{kuboformulas}) which provide response to a more physical situation.}.

\section{Conclusions} 
Supposing that the choice of thermopower is correct, the OCTHH
universality is partially vindicated by the model. The Kelvin and Kubo
formulae for the thermopower exhibit a sign change at $x_c(T)\approx
x_c(T=0)$, in parallel to the case in the cuprate
superconductors. Inherently, the model studied here is much too simple
to be taken seriously as a microscopic model of a real material: in
particular, it can do nothing to correlate optimal doping with a
QCP. It does, however, illustrate how a sign change in the thermopower
might ultimately be connected to a QCP. It also illustrates how an
equilibrium construction such as the Kelvin formula may be used to
approximate the behavior of a transport quantity such as the Kubo
formula.  Finally, we point out that the method used by
Vidyadhiraja, et al.~\cite{jarrell} to locate the quantum critical
point relies on the maximum in the entropy and hence is closely linked
with the Kelvin formula.  The fact that their state-of-the-art
calculations pinpoint optimal doping with a maximum in the entropy represents
an independent corroboration (albeit not an exact statement) that the
sign-change in the thermopower (that is the OCTHH universality) does
signify a QCP which has been further connected to the Mottness collapse~\cite{phillips,shila,phillips2}.

\acknowledgements
A. G. and B. S. S. were supported at UCSC by DOE under Grant No. FG02-06ER46319. 
P. W. P. would like to acknowledge partial support
from the NSF under Grant No. DMR-0940992 and the Center for Emergent
Superconductivity, an Energy Frontier Research Center funded by the
U.S. Department of Energy, Office of Science, Office of Basic Energy
Sciences under Award Number DE-AC0298CH1088.


\begin{thebibliography}{235}

\bibitem{kadowaki-woods} K. Kadowaki and S. B. Woods, Sol. St. Comm. {\bf 58}, 507 (1986).
\bibitem{heavy-fermi} N. Tsujii, H. Kontani, and K. Yoshimura, Phys. Rev. Letts. {\bf 94}, 057201 (2005). 
\bibitem{Tallon}
S. Obertelli, J.~R.~Cooper, and J.~L.~Tallon, Phys. Rev. B {\bf 46},
14928 (1992).
\bibitem{honmahor}T. Honma and P.H. Hor, Phys. Rev. B {\bf 77}, 184520 (2008).
\bibitem{presland}M. R. Presland, J. L. Tallon, R. G. Buckley, R. S. Liu, and N. E. Flower, Physica C {\bf 176}, 95 (1991).
\bibitem{false-hf}There is also discussion of universal behaviour of the thermopower for heavy fermion systems  of a rather different character from the one discussed by OCT. The relevant experiments are:
K. Behnia, D. Jaccard and J. Flouquet, J. Phys, C: Cond. Matt. {\bf 16}, 5187, (2004); J. Sakurai and Y. Isikawa, J. Phys. Soc. Japan {\bf 74}, 1926 (2005);
and theoretical discussion is in
 V. Zlatic, R. Monnier, J. K. Freericks, and K. W. Becker,   Phys. Rev. B 76, 085122 (2007).
\bibitem{leggett} A. J. Leggett, Nature Physics {\bf 2}, 134  (2006).
\bibitem{phillips} P. Phillips, T.-P. Choy, and R. G. Leigh, Rep. Prog. Phys. {\bf 72}, 036501 (2009).
\bibitem{shila}S. Chakraborty, D. Galanakis, and P. Phillips,
  Phys. Rev. B {\bf 82 }, 214503 (2010). 
\bibitem{Shastry}  M.~R.~Peterson, and B.~ S.~ Shastry, Phys. Rev. B {\bf 82}, 195105 (2010).
\bibitem{jarrell}N. S. Vidyadhiraja, A. Macridin, C. Sen, M. Jarrell, and M. Ma, Phys. Rev. Lett. {\bf
    102}, 206407 (2009).
\bibitem{Ashcroft-Mermin} Ashcroft N and Mermin N D 1976 Solid State Physics (Fort Worth, TX: Harcourt Brace Jovanovich College Publishers).
\bibitem{kelvin}W.~ Thomson  (Lord Kelvin)  Proc. R. Soc. Edinb. {\bf 123} (Collected Papers I, pp 237�41) (1854).
\bibitem{Shastry-review} B. ~S.~ Shastry, Rep. Prog. Phys. {\bf 72}, 016501 (2009).
\bibitem{Loram1}
J.~W.~Loram, K.~A.~Mirza,J.~R.~Cooper,W.~Y.~Liang, and J.~M.~Wade, J. Supercond. {\bf 7},243 (1994)
\bibitem{Loram2}
J.~W.~Loram,J.~Luo,J.~R.~Cooper,W.~Y.~Liang, and J.~L.Tallon, J. Phys. Chem. Solids, {\bf 62},59 (2001)  and Physica C {\bf 235-240}, 134 (1994).
\bibitem{Katsura}
S. ~Katsura, Phys. Rev. {\bf 127}, 1508 (1962).
\bibitem{Mccoy}
B. M. McCoy, E. Barouch and D. B. Abraham, Phys. Rev. A {\bf 4},2331 (1971).
\bibitem{Nambu}
Y. Nambu, Phys. Rev. {\bf 117}, 648 (1960).
\bibitem{phillips2}P. Phillips, Phil. Trans. A {\bf 369 }, 1574  (2011); see
  also, J. Zaanen and B. J. Overbosch, Phil. Trans. A {\bf 369 }, 1599  (2011).
\end{thebibliography}
\end{document}